\documentstyle[NATO,psfig]{crckapb} 
\def\deg{\hbox{$^\circ$}}

\def\fdg{\hbox{$.\!\!^\circ$}}  
\def\farcm{\hbox{$.\mkern-4mu^\prime$}}  
  
\def\NH{$N_{\rm HI}$~} 
\def \hi {H\,{\sc i~}} 
 
\def \Mg {Mg\,{\sc ii~}}
\def\kms{km\,s$^{-1}$} 
 
\def\vlsr{$v_{\rm LSR}$~}

\begin{opening}
\title{COMPACT HIGH--VELOCITY CLOUDS AT HIGH RESOLUTION}
\subtitle{}

\author{W.B. BURTON}
\institute{Sterrewacht Leiden\\
           P.O. Box 9513, 2300 RA Leiden, The Netherlands}
\author{R. BRAUN}
\institute{Netherlands Foundation for Research in Astronomy\\
           P.O. Box 2, 7990 AA Dwingeloo, The Netherlands}

\end{opening}

\runningtitle{COMPACT HIGH--VELOCITY CLOUDS}

\begin{document}

\begin{abstract}
Six examples of the compact, isolated high--velocity clouds catalogued by 
Braun \& Burton (1999) and identified with a dynamically cold ensemble of 
primitive objects falling towards the barycenter of the Local Group 
have been imaged with the Westerbork Synthesis Radio Telescope; an 
additional ten have been imaged with the Arecibo telescope.  The imaging 
reveals a characteristic core/halo morphology: one or several cores of 
cool, relatively high--column--density material, are embedded in an extended 
halo of warmer, lower--density material. Several of the cores show 
kinematic gradients consistent with rotation; these CHVCs are evidently 
rotationally supported and dark--matter dominated.  The imaging data allows 
several independent estimates of the distances to these objects, which 
lie in the range 0.3 to 1.0 Mpc.  The CHVC properties 
resemble what might be expected from very dark dwarf irregular galaxies. 

\end{abstract}

\begin{figure}[t] 
\psfig{figure=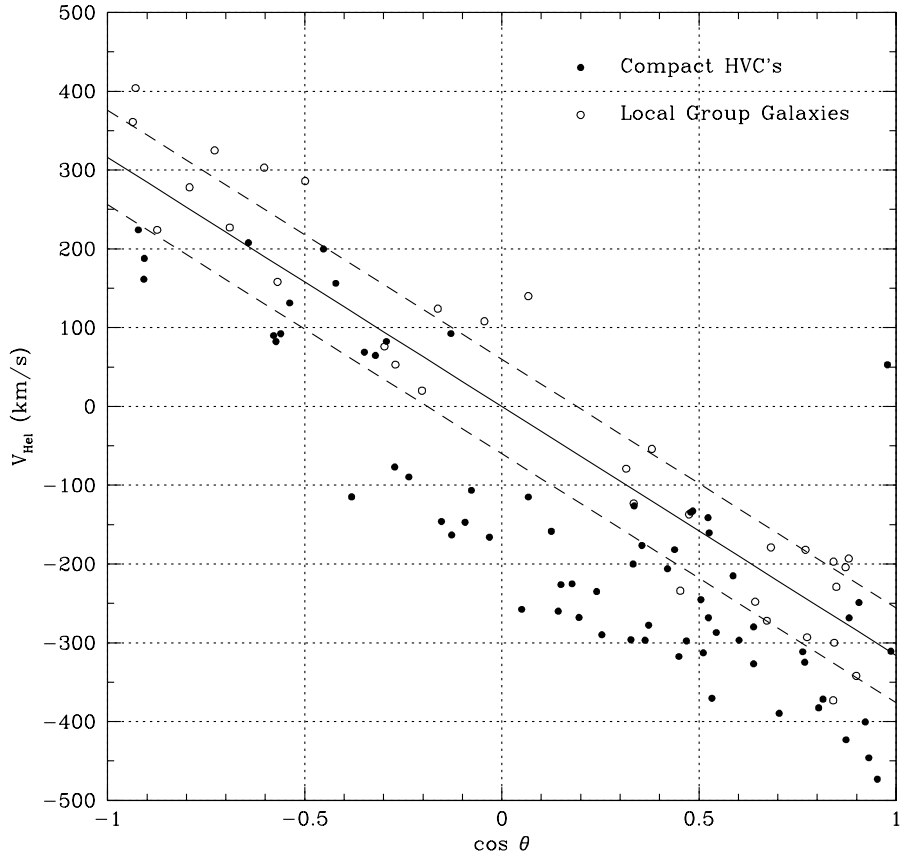,width=12.5cm}  
\vspace{-.6cm}  
\caption{Variation of heliocentric velocity versus the cosine of the 
angular distance between the solar apex and the $(l,b)$ direction of the 
compact, isolated high-velocity clouds catalogued by Braun \& Burton 
(1999).  The CHVCs are represented by filled circles; the galaxies 
constituting the Local Group, by open circles.  The solid line represents 
the solar motion of $v_\odot~=~316$~\kms~ toward $l=90\deg$, $b=-4\deg$ 
(Karachentsev \& Makarov 1996).  The dashed lines represent the envelope 
one standard deviation ($\pm 60$ \kms; Sandage 1986) about the 
velocity/angular--distance relation, pertaining for galaxies believed to be 
members of the Local Group. The kinematic and spatial deployment
of the CHVCs is consistent with that of a dynamically cold ensemble spread
throughout the Local Group, with a net infall velocity towards the barycenter
of the Local Group of some 100 \kms. 
 } 
\end{figure}

\section{Introduction}

     Hierarchical structure formation scenarios suggest that galaxies form 
by continuous accretion of small, dark--matter dominated satellites.  
The possibility of an extragalactic deployment of high--velocity clouds has 
long been considered, and in various contexts, by Oort (1966, 1970, 1981), 
Verschuur (1975), Eichler (1976), Einasto et al. (1976), Bajaja et al. 
(1987), Burton (1997), Wakker \& van Woerden (1997), 
Braun \& Burton (1999; 2000, astro-ph/9912417), Blitz et al. (1999), and 
L\'opez--Corredoira et al. (1999).  The discussion of Blitz et al. 
ties several HVC properties to the hierarchical structure formation and 
evolution of galaxies.  In this context the extended HVC complexes would be 
nearby and currently being accreted onto the Galaxy, while the compact, 
isolated objects would be the primitive building blocks at larger 
distances, scattered throughout the Local Group. 

     The class of compact, isolated high--velocity clouds catalogued by 
Braun \& Burton (1999) represent objects which plausibly originated under 
common circumstances, have shared a common evolutionary history, have not 
(yet) been strongly influenced by the radiative or tidal fields of the 
Milky Way or M31, and are falling towards the Local Group barycenter. The 
CHVC catalogue was based on survey data made  with telescopes of modest 
resolution.  The principal source was the Leiden/Dwingeloo Survey (LDS) of 
Hartman \& Burton (1997), characterized by the angular resolution of $36'$ 
provided by the Dwingeloo 25--meter antenna; important additional data came 
from the more coarsely--sampled surveys of Hulsbosch \& Wakker (1988) and 
of Bajaja et al. (1985) as analyzed by Wakker \& van Woerden (1991), as 
well as from some new \hi material observed at $21'$ resolution using the 
NRAO 140--foot telescope.  The CHVCs are largely unresolved in angle in the 
single--dish catalogue, and therefore the large range in observed velocity 
widths can not be directly interpreted in terms of the intrinsic 
properties of individual gaseous entities. 

     Of the sample of 65 CHVCs catalogued by Braun \& Burton (1999) only 
two had been subject earlier to interferometric imaging. Wakker \& Schwarz 
had used the WSRT to show that both CHVC\,$114\!-\!10\!-\!430$  and 
CHVC\,$114\!-\!06\!-\!466$ exhibit a core/halo structure (with only some 
40\% of the single--dish flux recovered) that the linewidths of the 
resolved cores were substantially narrower than when the individual cores 
were blended at low resolution, and that several of the components 
displayed systematic velocity gradients.  We have now imaged six 
additional CHVC fields using the Westerbork Synthesis Radio Telescope, and 
a further ten using the 304--meter Arecibo telescope.  Selected properties 
of several of these fields are shown here. A complete discussion of the 
WSRT imaging is given by Braun \& Burton (2000, astro-ph/9912417); a 
discussion of the Arecibo material is in preparation.

\section{WSRT and Arecibo Data}

     Observations of the six CHVC fields imaged with the WSRT involved 
twelve--hour integrations in the standard array configuration having a 
shortest baseline of 36 meters. The effective velocity resolution was 1.2 
times the channel 
spacing of 2.06 \kms, over 256 spectral channels centered on the \vlsr of
each source as catalogued by Braun \& Burton (1999) on the basis of 
the single--dish spectra in the LDS.  The angular and kinematic resolution
afforded by the WSRT makes it well suited in important regards to 
detailed studies of the CHVC class of objects.  Diffuse structures
extending over more than about 10 arcmin are, however, not adequately 
imaged by the interferometer unless precautions are taken to eliminate the 
short--spacing bowl surrounding regions of bright extended emission. In a
straightforward attempt to identify the column densities and overal extent
likely to characterize any diffuse structures, we made use of the LDS
data to determine the emission from an elliptical Gaussian with dimensions 
and orientation as measured 
in the LDS, and with a total flux sufficient to recover the LDS integrated
emission.  

\begin{figure}[t]  
\psfig{figure=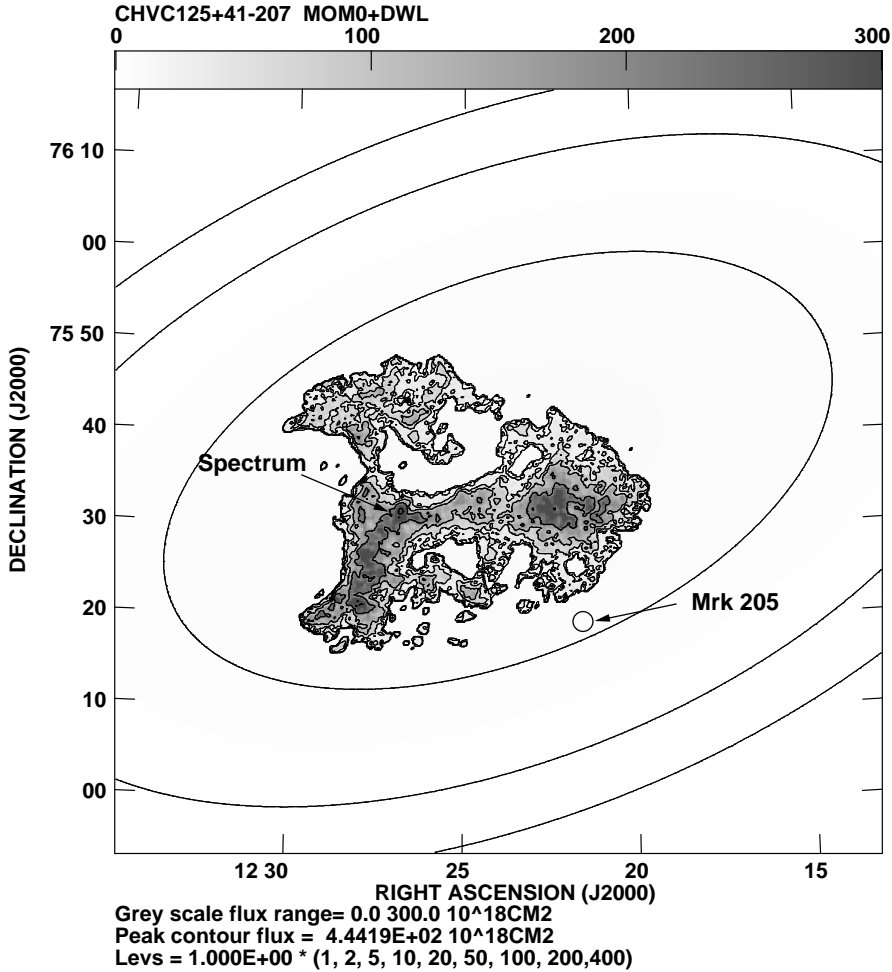,width=12cm}
\vspace{-.5cm}  
\caption{WSRT image of CHVC\,$125\!+\!41\!-\!207$ displaying \hi column 
densities at 28 arcsec angular resolution.  
\NH was calculated assuming negligible opacity, and is displayed by contours
at the levels indicated in units of $10^{18}$~cm$^{-2}$ and a linear 
grey--scale running from 0 to $300 \times 10^{18}$~cm$^{-2}$.
The location of the
Seyfert galaxy Mrk\,205 is marked.  This background source lies on a line 
of sight which penetrates the halo of the CHVC, where reconstruction 
of the integral flux using the composite WSRT and LDS single--dish 
data reveals 
a moderate diffuse--emission column depth.  Bowen \& Blades (1993) have 
measured \Mg absorption towards Mrk\,205, and a metallicity substantially 
subsolar.
 } 
\end{figure}  

\begin{figure}[t]  
\psfig{figure=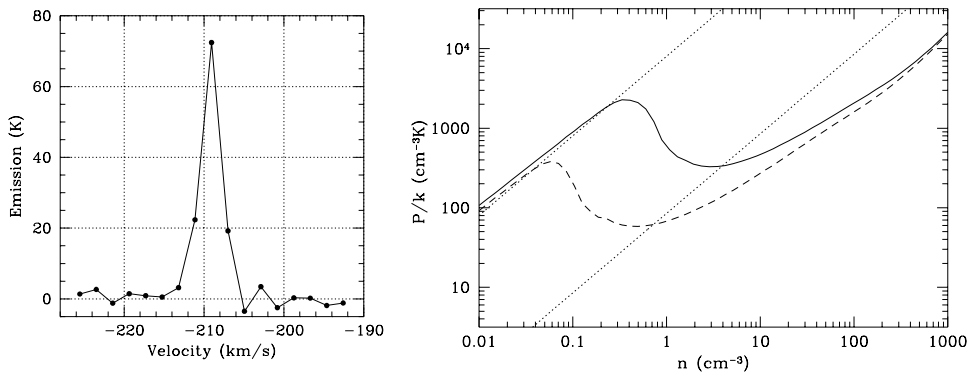,width=13cm}
\vspace{-.3cm}  
\caption{{\it left:}~ \hi spectrum observed in the direction indicated in 
Fig. 2 of one of the bright emission knots in CHVC\,$125\!+\!41\!-\!207$ .  
The spectrum is unresolved at a channel
separation of 2 \kms, indicating a core temperature of less than 85 K and
quiescent turbulence. {\it right:}~ Equilibrium temperature curves for \hi 
in an intergalactic
environment characterized by a metallicity of 10\% of the solar value and
a dust--to--gas ratio of 10\% of that in the solar neighborhood, calculated
for two values of the neutral shielding column depth, namely $10^{19}$~cm$^
{-2}$ (solid line) and $10^{20}$~cm$^{-2}$ (dashed line).  The upper dotted
line indicates the 8000 K temperature characteristic of the WNM; the lower
one, the kinetic temperature of 85 K observed in the opaque
core of CHVC\,$125\!+\!41\!-\!207$. The volume density is tightly 
constrained for this temperature; a distance then follows from the measured 
column density and angular size.
 } 
\end{figure} 

We have also recently observed ten CHVCs with the Arecibo telescope.
The Arecibo facility is well--suited to provide sensitivity to the
total column density at relatively high resolution.  This is especially
important for CHVC targets which are of a size comparable to the field
of view of most synthesis intruments.  The Arecibo targets were
observed with the new Gregorian feed and the narrow L--band receiver
with two bandpass settings, namely 6.25 MHz and 1.56 MHz (yielding
$\Delta v=1.3$ and 0.32 \kms, respectively) each centered on the \vlsr
of the CHVC as determined by Braun \& Burton (1999).  The Arecibo
targets were first mapped on a grid of $1\deg \times 1\deg$ size on a
fully--sampled 90 arcsec lattice, in short integrations, in order to
determine the locations of the peak flux concentrations; then at one or
more of these principal components long--integration spectra were
accumulated in a cut made at constant declination by repeating drift
scans over the same $2\deg$ in right ascension.  Typical column density
sensitivies of some $10^{17.5}$~cm$^{-2}$ over 20 \kms~were reached, an
\NH regime largely unexplored (cf. Zwaan et al. 1997).

\section{The Exceptionally Narrow Core in CHVC\,$125\!+\!41\!-\!207$}

     The compact high--velocity cloud CHVC\,$125\!+\!41\!-\!207$ is typical 
of the class of objects in several regards.  Figure 2, adopted from Braun 
\& Burton (2000, astro-ph/9912417; 
see also Burton \& Braun 1999) shows several cool, quiescent cores embedded 
in a diffuse, warmer halo. The spectrum plotted in Fig. 3, observed towards 
the brightest of these 
cores, has a linewidth which is completely unresolved by the effective 
resolution of the WSRT imaging. The velocity channels adjacent to the line
peak have intensities below 20\% of the maximum value.  Such a
width is one of the narrowest measured in \hi emission, and contrains
both the kinetic temperature and the amount of turbulence.  An upper limit
to the thermal--broadening FWHM of 2 \kms~corresponds to an upper limit to
the kinetic temperature of 85 K.  The physical situation is yet more 
tightly
constrained, because the brightness temperature in this core is observed to
be 75 K; thus a lower limit to the opacity follows from $T_{\rm b} =
T_{\rm k}(1 - e^{-1})$, yielding $\tau \geq 2$. Any broadening which might
be due to macroscopic turbulence is less than 1 \kms.

     The tightly--constrained temperature found for 
CHVC\,$125\!+\!41\!-\!207$ allows an estimate of the distance to this 
object.  Wolfire et al. (1995a,
1995b) show that a cool \hi phase is stable under extragalactic conditions
if a sufficient column of shielding gas is present and if the thermal
pressure is high.  Calculations of equilibrium conditions which would pertain
in the Local Group environment have been communicated to us by Wolfire,
Sternberg, Hollenbach, and McKee, and are shown in Fig. 3 for two 
bracketing
values of the shielding column density, namely $10^{19}$~cm$^{-2}$ and 
$10^{20}$~cm$^{-2}$.  The figure shows that the equilibrium
volume densities corresponding to the observed value of $T_{\rm k}=85$~K lie
in the range 0.65 to 3.5~cm$^{-3}$.  Thus provided with this range of 
volume
densities, and having measured both the column depth of the cool core and
its angular size, the distance to CHVC\,$125\!+\!41\!-\!207$ follows 
directly
from $D=N_{\rm HI}/(n_{\rm HI}/\theta)$, yielding a value in the range
210 to 1100 kpc.  Further considerations of several opaque, cool cores made
by Braun \& Burton (2000, astro-ph/9912417) suggest
that a distances between 0.5 and 1 Mpc are most plausibly representative of
these objects.

     Measurements of metallicity of high--velocity clouds are important to 
discussions of the nature of the phenomenon.  If the clouds are primitive
objects scattered throughout the Local Group, the gas would not be 
substantially enriched in heavy elements produced by stellar evolution.  On
the other hand, if the anomalous velocities had been generated by supernova
explosions or some other energetic occurance in the Galactic disk, for 
example according to the precepts of the galactic fountain scenario 
(Shapiro \& Field, 1976; Bregman, 1980) then
the gas would be substantially enriched, with the already moderately high 
metallicities characteristic of the Galactic disk further enhanced
by the circumstances of the ejection event.  The small angular size of 
the CHVCs, and the amount of substructure being revealed at high 
resolution, will make it generally difficult to find suitable background 
sources.  But the diffuse halo of CHVC\,$125\!+\!41\!-\!207$  overlaps the 
Seyfert galaxy Mrk\,205, and in this direction Bowen \& Blades (1993) 
detected \Mg absorption at $v_{\rm LSR}=-209$~\kms.  We determine a 
metallicity of this object in the range 0.04 to 0.07 solar.

\section{Rotation in the Cores of CHVC\,$204\!+\!30\!+\!075$}

\begin{figure}[t]  
\psfig{figure=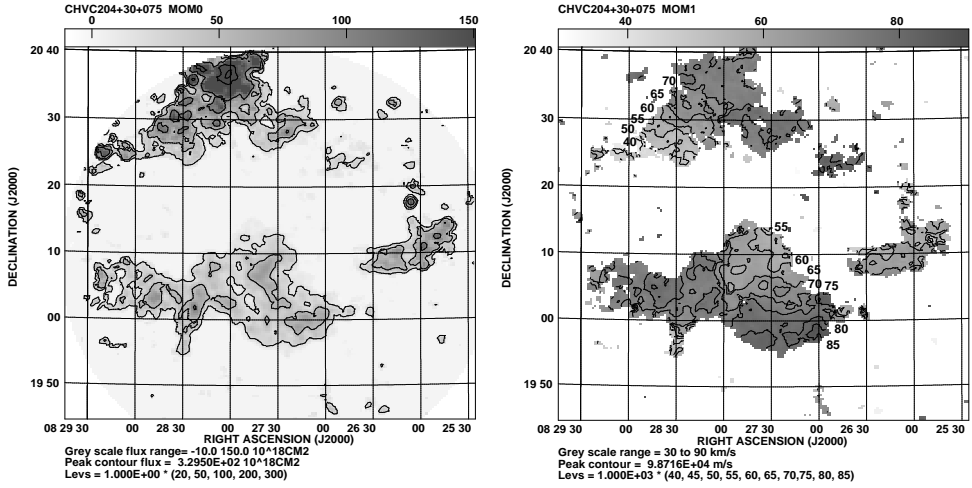,width=13cm}
\caption{{\it left:}~ Westerbork image of CHVC\,$204\!+\!30\!+\!075$ 
showing \NH (calculated assuming negligible opacity) at an angular 
resolution of 
1 arcmin; contours are drawn at levels of 20, 50, 100, 200, and $300 \times
10^{18}$~cm$^{-2}$.  {\it right:}~ Intensity--weighted 
line--of--sight velocity, with contours of $v_{\rm LSR}$ showing systematic 
kinematic gradients across the two principal components of the CHVC object, 
consistent with rotation; contours are drawn in steps of 5 \kms~from 40 to 
85 \kms.  } 
\end{figure}

\begin{figure}[t]  
\psfig{figure=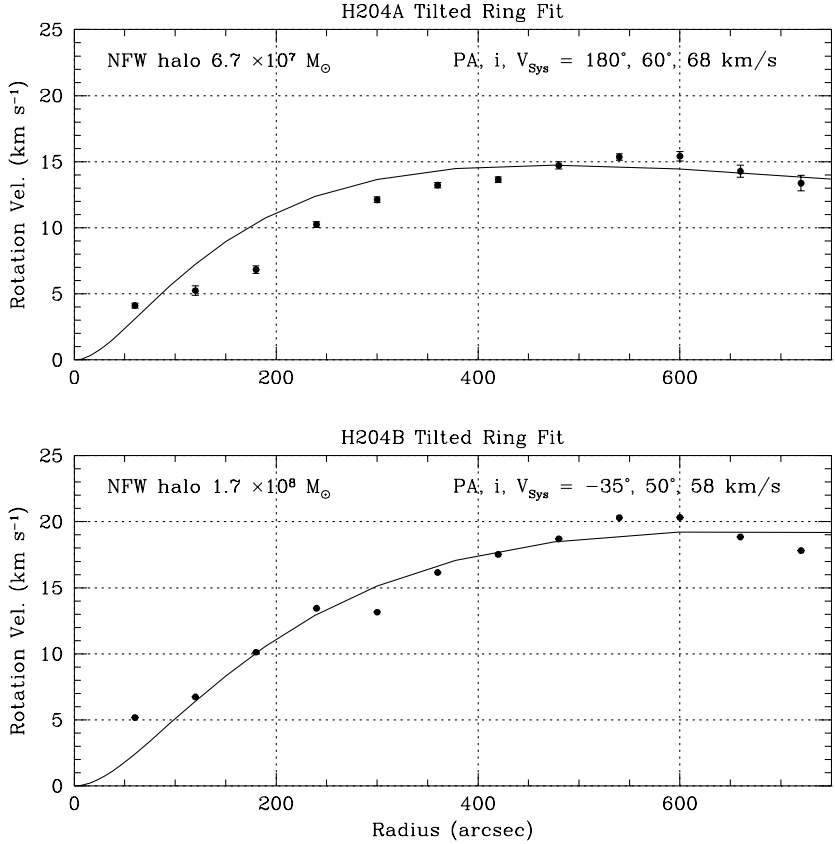,width=12cm}
\vspace{-.5cm}  
\caption{Rotation velocities fit by a standard application of the 
tilted--ring method to kinematic gradients revealed in the two 
principal components of 
CHVC\,$204\!+\!30\!+\!075$ shown in Fig. 4.  (The upper panel pertains to 
the component at higher declination.) The best--fit position angle, 
inclination,
and systemic velocity are indicated.  The solid lines show the rotation 
curves of Navarro, Frenk, \& White (1997) cold--dark--matter halos of the
indicated masses.
}
\end{figure}  

     The narrowest FWHM of the CHVCs cataloged by Braun \& Burton had a 
value of 5.9 \kms; the broadest, a value of 95 \kms.  Under higher 
resolution, the characteristic width narrows as objects are resolved into 
several principal components, moving relative to each other.  
If the objects with multiple cores are to be stable, distances of order 
several hundred kpc are required.  Some of the compact cores imaged owe 
their large total width in the LDS single--dish data to velocity 
gradients.  The resolved WSRT image of 
CHVC\,$204\!+\!30\!+\!075$ is shown in Fig. 4.  The object shows two 
principal components each of which is elongated; furthermore, each of the 
elongated structures shows a systematic velocity gradient along the major 
axis.  

     The velocity gradients exhibited by the two principal components of 
CHVC\,$204\!+\!30\!+\!075$ can be modelled in terms of circular rotation in 
a flattened disk system.  Fig. 5 shows the results of fitting a standard 
tilted--ring model to the data.  The fits display velocity rising slowly 
but continuously with radius to an amplitude of some 15 \kms~in the one 
case and to some 20 \kms~in the other, and then flattening to a constant 
value beyond about 500 or 600 arcsec.  Estimates of the contained dynamical 
mass follow from the rotation curves if the distance is assumed, and the total
gas mass follows from the integrated \hi fluxes.  At an assumed distance of 
0.7 Mpc, the two principal clumps of CHVC\,$204\!+\!30\!+\!075$ have 
$M_{\rm dyn}=10^{8.1}$ and $10^{8.3}$  M$_\odot$, and gas masses 
(including \hi and helium of 40\% by mass) of $10^{6.5}$ and $10^{6.9}$ 
M$_\odot$, respectively for the upper and lower concentrations shown in 
Figures 4 and 5. The dark--to--visible mass ratios for these concentrations 
are 36 and 29, respectively.  

     The shape of the modelled rotation curves for both of the 
CHVC\,$204\!+\!30\!+\!075$ components is reproduced by the standard 
cold--dark--matter halo as presented by Navarro et al. (1997).  At the assumed 
distance of 0.7 Mpc, the Navarro et al. halos fit to the two components 
have masses of $10^{7.8}$ M$_\odot$ (within 9.3 kpc) and $10^{8.2}$ M$_\odot$ 
(within 12.6 kpc), respectively.

\section{The Objects CHVC\,$230\!+\!61\!+\!165$ and 
CHVC\,$158\!-\!39\!-\!285$} 

 
\begin{figure}[t]  
\psfig{figure=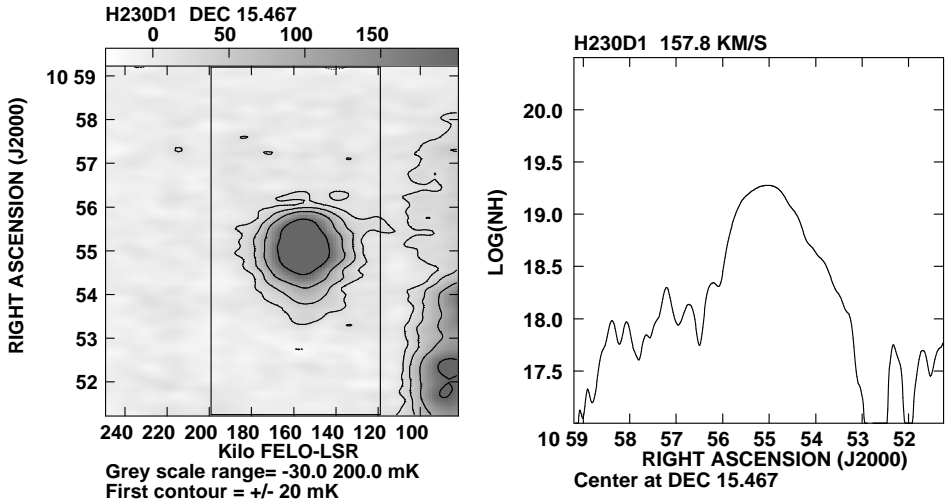,width=12.8cm}
\vspace{-.5cm}  
\caption{{\it left:}~ Position, velocity cut through 
CHVC\,$230\!+\!61\!+\!165$ 
observed with the Arecibo telescope, at an angular resolution of 
$3\farcm3$. The cut samples right ascension along the
fixed declination $15\fdg467$.  This compact object shows no kinematic 
gradient along the cut sampled. {\it right:}~ Variation of \NH with 
position, sampled at the velocity (158 \kms) of the peak of the 
column--density distribution. The \NH values plotted here and in Fig. 7 
are based on 
calibrated intensities in units of $T_{\rm b}$.
 } 
\end{figure}  

\begin{figure}[]  
\psfig{figure=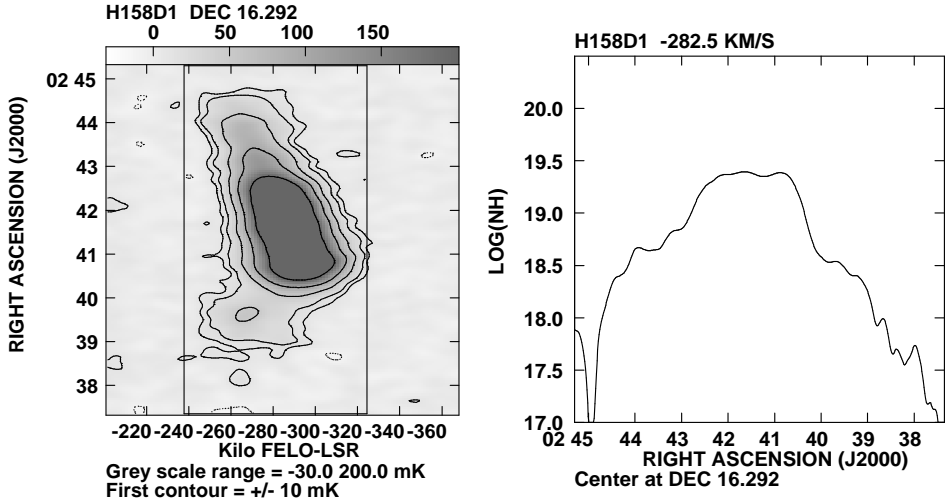,width=12.8cm}
\vspace{-.5cm}  
\vfill  
\caption{{\it left:}~ Position, velocity cut through  
CHVC\,$158\!-\!39\!-\!285$ observed with the Arecibo telescope at an 
angular resolution of $3\farcm3$.
The cut samples right ascension along the
fixed declination $16\fdg292$. This CHVC shows a kinematic gradient 
consistent with rotation, spanning 
some 40 \kms, as well as the characteristic core/halo morphology. {\it 
right:}~ Variation of \NH
with position, sampled at the velocity ($-282.5$ \kms) of the peak of the
column--density distribution.  
 } 
\end{figure} 

The WSRT imaging of CHVC\,$230\!+\!61\!+\!165$ revealed a simple, faint
structure. The Arecibo telescope is particularly well-suited to such
targets, because of its sensitivity to low \hi brightnesses.  A
position, velocity cut through this object at the location of the peak
\NH is shown in the lefthand panel of Fig. 6.  No kinematic gradient is
revealed.  However the cut does show an interesting characteristic
which several other of the CHVCs observed at Arecibo also show, namely
a tendency to be more sharply bounded on one side of the cut than on
the other.  In the case of CHVC\,$230\!+\!61\!+\!165$, the higher
right--ascension boundary is sharper than the lower one down to \NH =
10$^{18.5}$.  (We will consider this property further in our full
discussion of the Arecibo observations.)

    The Arecibo data on CHVC\,$158\!-\!39\!-\!285$ are shown in Fig. 7. 
This object is also a simple one, with only one component revealed.  The
position, velocity cut through the location of the peak \NH value also shows
that one side of the object is more sharply bounded than the other.  The
prominent kinematic gradient is consistent with simple rotation within
the high \NH core.  The plots
on the righthand panels of Fig. 6 and Fig. 7 show the variation of \NH
with position across the two CHVCs.  The cores are embedded in diffuse 
material with characteristic \NH values of order $10^{18.5}$ cm$^{-2}$.

\section{Discussion}

The WSRT and Arecibo high--resolution imaging shows that the morphology
of the compact high--velocity clouds is characterized by one or more
quiescent, low--dispersion compact cores embedded in a diffuse, warmer
halo.  The gas in the cores is identified with the cool neutral medium
(CNM) of condensed \hi at temperatures in the range 50 -- 200 K; the
halo gas is identified with the warm neutral medium.  Such a nested
geometry is expected if the CNM is to be stable in the presence of an
ionizing radiation field of the sort expected in the Local Group
environment. The cores contribute typically about 40\% of the \hi flux,
while covering about 15\% of the surface in the CHVC.

The high--resolution imaging data allows two independent distance
estimates to be made.  A distance for CHVC\,$125\!+\!41\!-\!207$
follows from assuming rough spherical symmetry and equating the
well--constrained volume and column densities of the compact cores.
Another distance constraint (coupled with a dark--to--visible mass
ratio) follows from consideration of the stability of CHVCs having
multiple cores in a common envelope but having large relative
velocities.  The evidence available indicates that CHVCs have
characteristic sizes of about 10 kpc, \hi masses of about $10^7$
M$_\odot$, and are at distances of 0.4 -- 1.0 Mpc.

The imaging has also shown that the compact cores of CHVCs are commonly
rotating.  The observed kinematic gradients can be fit by rotating
flattened disks, yielding dynamical masses for assumed distances.
Dark--to--visible mass ratios of order 10 -- 40 at $D=0.7$~Mpc are
indicated.  The rotation curves shown here agree well with the
cold--dark--matter halo predicted by Navarro et al. (1997).

The gaseous properties characteristic of CHVCs bear similarities to
those of many dwarf irregular galaxies populating the Local Group (e.g.
Young \& Lo 1996, 1997a, 1997b).  It is an important astrophysical
challenge to establish the details of these similarities.  It is
particularly important to establish whether any of the CHVCs contain
stars as well; the stellar density might be very low, but detection of
any stellar component would lead to improved distances and would
constrain the evolutionary history, and would augment the information
on the faint end of the Local Group luminosity function.  Failure to
detect stars would imply that CHVCs are very primitive proto--galactic
objects dominated by dark--matter halos.  In either case, it seems
plausible that the CHVCs are the missing Local Group satellite systems
predicted by Klypin et al. (1999) and Moore et al. (1999).

\acknowledgements
We are grateful to M.G. Wolfire, A. Sternberg, D. Hollenbach, and C.F. McKee
for providing the equilibrium temperature curves shown in Fig. 3, and to 
P. Perillat for assistance during our Arecibo observing session.  
The WSRT is operated by the Netherlands Foundation for Research in 
Astronomy, under contract with the Netherlands Organization for 
Scientific Research.  The Arecibo Observatory is part of the National 
Astronomy and Ionosphere Center, which is operated by Cornell University 
under a cooperative agreement with the National Science Foundation.

\end{document}